\documentclass[conference]{IEEEtran}
\IEEEoverridecommandlockouts
\usepackage{cite}
\usepackage{amsmath,amssymb,amsfonts}
\usepackage{algorithmic}
\usepackage{graphicx}
\usepackage{textcomp}
\usepackage{xcolor}
\usepackage{bm}

\usepackage{circuitikz}
\usepackage{tikz}
\usepackage{standalone}
\usetikzlibrary{external}
\usetikzlibrary{calc}

\def\BibTeX{{\rm B\kern-.05em{\sc i\kern-.025em b}\kern-.08em
    T\kern-.1667em\lower.7ex\hbox{E}\kern-.125emX}}
\begin{document}

\title{Priority Lists for Power System Investments: \\ Locating Phasor Measurement Units\\
% \thanks{\hrule \vspace{.1cm}
% This work was partially funded by the Skolkovo Institute of Science and Technology as a part of the Skoltech NGP Program (Skoltech-MIT joint project).}
% This work is supported by the ID-EDGe project, funded by Innovation Fund Denmark, Grant Agreement No. 8127-00017B, and by the Skolkovo Institute of Science and Technology as a part of the Skoltech NGP Program (Skoltech-MIT joint project).}
}

\author{\IEEEauthorblockN{Ilgiz Murzakhanov}
\IEEEauthorblockA{\textit{Department of Electrical Engineering} \\
\textit{Technical University of Denmark}\\
Kgs. Lyngby, Denmark \\
ilgmu@elektro.dtu.dk}
\and
\IEEEauthorblockN{David Pozo}
\IEEEauthorblockA{\textit{Center for Energy Science and Technology} \\
\textit{Skolkovo Institute of Science and Technology}\\
Moscow, Russia \\
d.pozo@skoltech.ru}
% \and
% \and
% \and
% \and
}

\maketitle
\IEEEpubidadjcol
\begin{abstract}
Power systems incrementally and continuously upgrade their components, such as transmission lines, reactive capacitors, or generating units. Decision-making tools often support the selection of the best set of components to upgrade. Optimization models are often used to support decision making at a given point in time. After certain time intervals, re-optimization is performed to find new components to add. 
In this paper, we propose a decision-making framework for incrementally updating power system components.  This is an alternative approach to the classical sequential re-optimization decision making for an investment problem with modeled budget constraints. Our approach provides a priority list as a solution with a list of new components to upgrade. We show that i) our framework is consistent with the evolution of power system upgrades, and ii) in particular circumstances, both frameworks provide the same solution if the problem satisfies submodularity property.

We have selected the problem of phasor measurement unit localization and compared the solution with the classical sequential re-optimization framework. For this particular problem, we show that the two approaches provide close results, while only our proposed algorithm is applicable in practice. The cases of 14 and 118 IEEE buses are used to illustrate the proposed methodology.

%Power systems are incrementally and continuously updating the elements that constitute them, such as transmission lines, reactive capacitors, or generator units. Decision-making tools usually support the selection of the best set of components to upgrade. Optimization models are typically used to support decision-making at a particular moment in time. It is required to calculate new solutions in fixed time intervals by solving an optimization model with the latest data and an updated list of components. In this paper, we propose a decision-making framework to upgrade power system components incrementally.  It is an alternative approach to the classical sequential re-optimization decision-making for an investment problem with modeled budget constraints. Our approach provides a  priority list as a solution with a list of new components to upgrade. In this paper, we show that i) our framework is consistent with the evolution of the power system upgrades, and ii) in particular circumstances, both frameworks provide the same solution. We have selected the problem of locating phasor measurement units (PMUs) and compared the solution with the classical sequential re-optimization framework. For this particular problem, we show that the two approaches provide close results, while only our proposed algorithm is practically applicable. An IEEE 14-bus and 118-bus cases are used to illustrate our proposed methodology.
\end{abstract}


\section{Introduction}
The evolution of the power system requires continuous updating of its components to meet requirements of reliability, sustainability, and economic efficiency. Phasor measurement units (PMUs) are becoming one of the central and most promising components of the modern power system. The installation of PMUs in power systems can solve two problems: provide observability of the system and increase measurement accuracy. The latter is achieved by the property of PMU measurements, which are time-synchronized and sampled at 30-60 times a second \cite{AntonioJ.Conejo2018}. However, the installation and maintenance costs of PMUs are not negligible. Thus, a trade-off between the achieved accuracy of the measurements and the incurred cost must be sought during placement. For this reason, the installation PMUs is done in multiple stages, when at each stage, the best location of PMUs is defined for the given budget. In this paper, we present a methodology for building a priority list for the multi-stage PMU placement problem.

\subsection{Illustrative Sequential Investment Example}
To motivate to the general mathematical formulation of sequential multi-stage investment problems, we draw a parallel illustrative example based on the well-known knapsack problem. 
In the original knapsack problem, a set of items with known weights and values is given. We aim to pick the items with a maximum total value while the total weight must not exceed a set limit. As an illustrative example, we use the following optimization problem:
\begin{subequations}
\begin{alignat}{2}
\max_{x_1,x_2,x_3,x_4} \quad & 7x_1+5x_2+4x_3+x_4\label{ex_obj} \\
\mbox{s.t.}\quad
& 4x_1+2x_2+3x_3+6x_4 \leq b   \label{ex_b} \\
% & \sum^{4}_{i=1} x_i = c \label{ex_sum} \\
& x \in \{0,1\}   \label{ex_bin}
\end{alignat}
\end{subequations}

The objective is representing the value provided by the optimal selection among four items \eqref{ex_obj}. The budget of the optimization problem (\ref{ex_obj})-(\ref{ex_bin}) is defined by $b$. Equation (\ref{ex_b}) represents the budget limit related to the weights of the selected items. In the context of power systems investments, (\ref{ex_b}) is typically representing a limited budget (e.g., monetary budget) for investment decisions. 

This particular example can be easily solved for different levels of $b$, i.e., different budgets. The optimal solutions to this problem are presented in Table \ref{tab_budg}. We observe that during an increase of $b$ from the interval $[2,4)$ to the interval $[4,6)$ there is a change of the optimal solution from $x_2$ to $x_1$. It shows that sequential optimization could lead to sub-optimal solutions in multi-stage decision-making. Therein, we refer to this method as \textit{optimal budget-constrained method}.
\begin{table}[htbp]
\caption{Solution of the optimal budget-constrained method}
\begin{center}
\begin{tabular}{|c|c|c|}
\hline
\textbf{\bm{$b$} value in (\ref{ex_b})}& \textbf{Optimal solutions(s)}& \textbf{Objective value in (\ref{ex_obj})} \\
\hline
[0,2) & - & 0 \\
\hline
[2,4) & $x_2$ & 5 \\
\hline
[4,6) & $x_1$ & 7 \\
\hline
[6,9) & $x_1$, $x_2$ & 12 \\
\hline
[9,15) & $x_1$, $x_2$, $x_3$ & 16 \\
\hline
[15,$+\infty$) & $x_1$, $x_2$, $x_3$, $x_4$ & 17 \\
\hline
\end{tabular}
\label{tab_budg}
\end{center}
\end{table}

%In principle, there are two distinct approaches of tackling the problem (\ref{ex_obj})-(\ref{ex_bin}). The first approach is the sequential re-optimization decision-making method for an investment problem with modeled budget constraints. For compactness of the formulation, this approach is further referred to as the budget-based method. The idea of the approach is at each stage to choose the best options; even that would require to undo the previous choices. While the second approach 

An alternative to the optimal budget-constrained method, we can use an approach that seeks the best option at each stage while maintaining the solutions made in previous steps. As a result, the choices made in the earlier stages are always included in the solution of the next stages. Further, we refer to this second approach as the \textit{greedy method} (for its simplicity). 

The solutions of the greedy method is given in Table  \ref{tab_greed}. 
Based on the obtained solutions and their mutual comparison, we draw the following conclusions. First, during an increase of $b$ from 2 to 5 in the optimal budget-constrained method, there is a change of an optimal variable from $x_2$ to $x_1$ as it allows to maximize the objective. On the contrary, in the greedy algorithm, the next solutions always include the previous decisions on the optimal variables. Second, both methods for the given problem (\ref{ex_obj})--(\ref{ex_bin}) produce the same results on the intervals $b \in [0,4)$ and $b \in [9,+\infty)$. Third, on the interval $b \in [4,9)$, the optimal budget-constrained method obtains a higher objective for the maximization problem. 
\begin{table}[htbp]
\caption{Solution of the greedy method}
\begin{center}
\begin{tabular}{|c|c|c|}
\hline
\textbf{\bm{$b$} value in (\ref{ex_b})}& \textbf{Solution(s)}& \textbf{Objective value in (\ref{ex_obj})} \\
\hline
[0,2) & - & 0 \\
\hline
[2,5) & $x_2$ & 5 \\
\hline
[5,9) & $x_2$, $x_3$ & 9 \\
\hline
[9,15) & $x_2$, $x_3$, $x_1$ & 16 \\
\hline
[15,$+\infty$) & $x_2$, $x_3$, $x_1$, $x_4$ & 17 \\
\hline
\end{tabular}
\label{tab_greed}
\end{center}
\end{table}

From the illustrative case study presented, we resolve that \textit{
while the first approach always provides the optimal solution for a given budget $b$, the second approach provides a sorted list of irrevocable decisions to make, i.e., a priority list that in principle is not optimal for the budget level $b$.  
We will show that in particular problems that have submodularity property, both approaches provide the same solution. }
Next, we provide a literature review on the application of sequential decision-making in power systems.

\subsection{Literature Review}
New components' installation and in power grids is done periodically and is tightly connected with the optimal investment problem \cite{farrokhifar2020energy}. In \cite{Moreno2009}, the authors propose a scheme for facilitating the efficient and timely expansion of transmission capacity. The scheme is suitable for markets with high demand growth and a high probability of increasing power flows in the transmission system. In the proposed solution, the anticipatory investment allows avoiding duplicating the transmission lines and a sharp increase of reinforcing cost, as the lines were not initially designed for future upgrade. 
%However, according to numerical simulations, the proposal is not robust. 
In \cite{Zhang2018}, the multi-stage transmission expansion problem is extended by considering \mbox{N-1} security constraints. The problem is addressed by applying a reduced disjunctive model (RDM) with exact linearization of \mbox{N-1} contingency constraints. Compared with traditional approaches, the proposed solution needs fewer binary variables so that the computational performance is greatly improved. A two-stage robust generation expansion problem is presented in \cite{Dehghan2014}. The proposed solution determines the optimal generation mix and construction time of new power plants to minimize the total discounted utility costs. The authors present a two-stage robust optimization methodology with uncertain investment and operation costs and uncertain electricity load demands. An integrated transmission and generation planning problem in a restructured market environment proposed in  \cite{Jin2014}.
%and \cite{pozo2012three}.
In this work, the coordination of new generation capacity investments and transmission lines is modeled using a tri-level optimization model.
%with a centralized expansion planning problem at the top level, responses of multiple transmission companies on their generation expansion decisions at the middle level, and a market-clearing problem at the bottom level. 
However, the proposed models are static and consider only a single representative period in a future year. To avoid this issue, other works have been including several representative operating points, in the context of transmission expansion planning under uncertainty, see works on distributionally robust optimization \cite{velloso2020distributionally} and in the context of distribution resource planning \cite{lopez2021empirical}. 
However, the above models are static and typically only consider a single representative period in a future target year. To circumvent this shortcoming, other studies have been including several representative operating points, in the context of transmission expansion planning under uncertainty, see the works on distributionally robust transmission planning  \cite{velloso2020distributionally},  work in climate-aware transmission planning \cite{moreira2021climate}, and research on  distribution grid resource planning \cite{lopez2021empirical}. 
Other works have been considering the joint investment of storage devices and transmission lines \cite{bustos2017energy}.  
%In \cite{Chu2017}, the authors introduce an optimal placement method for dynamic power compensation devices for improving the immunity to communication failures in multi-infeed high-voltage direct current (HVDC) systems.

% The problem of sensor placement is explored in different domains: general dynamical networks, traffic and power systems. 
The optimal sensor and actuator placement in dynamical networks is presented in \cite{Summers2014}. The paper introduces mathematical terms of controllability and observability of the networks, definition of the modular function, and their applicability for optimal sensor placement problem. The authors introduce a linear dynamical system and prove that a described algorithm obtains an optimal solution under preserved modularity properties. The idea of modular functions is further explored in \cite{Summers2016}. The authors show that several important classes of metrics based on the controllability and observability Gramians
have a strong structural property. This property provides either an efficient global optimization or an approximation guarantee by using a simple greedy heuristic. 
% A similar conclusion is obtained in \cite{Mehr2018}, where a submodular approach is applied for optimal sensor placement in traffic. While deploying a large number of sensors leads to superior performance, the installation and maintenance costs of sensors are high. As a result, the authors seek a trade-off between system performance and the incurred cost. The work provides mathematical proof that a problem of optimal sensor placement in traffic networks has a submodularity property. Therefore, for its efficient solution, the greedy algorithm can be utilized. 
% Determination of optimal PMU placement for fault-location observability is explored in \cite{Mazlumi2008}. The proposal forms the binary optimization problem and utilizes the branch and bound method to determine the minimum number of PMUs for accurately finding the location of any fault in the power system. However, the introduced solution cannot be extended for a multi-stage PMU installation problem.
Determination of optimal PMU placement for fault-location observability is explored in \cite{5357471}. The proposal first formulates the conventional complete observability of power networks and then adds different contingency conditions in power networks, including measurement losses and line outages. However, the introduced solution is not extended for a multi-stage PMU installation problem.
In \cite{Dua2008}, optimal multi-stage scheduling of PMU placement is investigated. The authors utilize integer linear programming for improving the observability of the power system in terms of two metrics: bus observability index and system observability redundancy. Optimal selection of PMU locations for enhanced state estimation in smart grids is explored in \cite{Rihan2013}. While the paper presents a few new ideas, it does not consider the multi-stage installation of PMUs.

\subsection{Paper Contributions and Organization}
The contributions of this work are the following:
\begin{itemize}
    \item We formulate the multi-stage PMU placement problem for increasing measurement accuracy with budget constraints, which was not done before.
    \item We compare several metrics for assessing the accuracy of PMU measurements and provide recommendations for their use. 
    \item We prove that the problem of PMU location for increasing measurement accuracy does not have a submodular structure.
    \item We compare the performance of optimal budget-constrained and greedy algorithms to PMU placement problem on IEEE 14-bus and 118-bus cases.
\end{itemize}

% \subsection{Outline}
The remainder of this paper is organized as follows. First, necessary notations and background on submodularity and measurement accuracy are given in section~\ref{NotBack}. In section \ref{PropMet}, we describe the followed methodology. The numerical results of all simulations are provided in section~\ref{NumRes}. Finally, section~\ref{Conc} concludes the paper and proposes future work.

\section{Preliminary Mathematical Background}\label{NotBack}
In this section, we explain the concept of submodularity and the metrics for evaluation of PMU measurement accuracy, which are used further for the PMU location problem.

\subsection{Submodularity}
By conventional notation, the set of real numbers is denoted $\mathbb{R}$. For a finite set $\mathcal{C}$, $|\mathcal{C}|$ is its cardinality. The set of all subsets of $\mathcal{C}$ is denoted by $2^{\mathcal{C}}$. Consider the set functions of the form: $f : 2^{\mathcal{C}} \rightarrow \mathbb{R}$. 

\noindent\textbf{Definition 1.} \emph{A set function $f : 2^{\mathcal{C}} \rightarrow \mathbb{R}$ is called monotone decreasing if for all} $A, B \subseteq \mathcal{C}$, \emph{the following is true:}
\begin{equation} \label{monotone}
A \subseteq B \textrm{   \emph{if and only if }} f(A) \geq f(B)
\end{equation}

\noindent\textbf{Definition 2.} \emph{A set function} $f : 2^{\mathcal{C}} \rightarrow \mathbb{R}$ \emph{is called submodular if for all subsets} $A \subseteq B \subseteq \mathcal{C}$ \emph{and all elements} $s \notin B$, \emph{it holds that:}
\begin{equation} \label{submodular}
f(A \cup {s}) - f(A) \geq f(B \cup {s}) - f(B) 
\end{equation}

% \emph{or equivalently, if for all subsets} $A, B \subseteq \mathcal{C}$, \emph{it holds that:}

% \begin{equation} \label{submodular2}
% f(A) + f(B) \geq f(A \cup B) + f(A \cap B)
% \end{equation}

The interpretation of (\ref{submodular}) follows from a property of diminishing returns: addition of an element $s$ to a smaller set, which is a subset of the larger set, will lead to a larger increase in $f$ \cite{lovasz1983submodular}. A set function is called \emph{supermodular} if the reversed inequality in (\ref{submodular})  holds.

% A following theorem is a powerful tool in proving submodularity of set functions \cite{Mehr2018}.

\noindent\textbf{Theorem 1}  \cite{lovasz1983submodular}\textbf{.} \emph{A set function} $f: 2^{\mathcal{C}} \rightarrow \mathbb{R}$ \emph{is submodular if and only if for a set} $Q$, \emph{the set functions} $f_{s}: 2^{Q-s} \rightarrow \mathbb{R}$ \emph{defined as} $f_{s}(Q) = f(Q \cup s) - f(Q)$ \emph{are monotone decreasing} $\forall s \in \mathcal{C}$.

\noindent\textbf{Observation on optimization of submodular functions.} \emph{Given the maximization problem of a submodular function $f$ with the following form:}
\begin{subequations}
\begin{alignat}{2}
\max_{Q \subseteq \mathcal{C}} f(Q)  \label{obj}\\
\mbox{s.t.}\quad
 |Q| &\leq k   \label{constr}
\end{alignat}
\end{subequations}
\emph{where $k$ is the maximum allowed cardinality. For submodular function $f$, optimization problems of the form (\ref{obj})-(\ref{constr}) can be solved in polynomial time with certain optimality bounds with the use of greedy algorithms}. 

For the detailed formulation of the greedy algorithm, we refer to \cite{Mehr2018}. The main idea of the greedy algorithm is at each iteration to simply add the element which maximizes the gain of $f$. Further, in section \ref{NumRes} we check if multi-stage PMU location problem has submodular structure and implement two algorithms.

\subsection{Measurement Accuracy}\label{AA}

Next, we present the weighted least squares (WLS) method in phasor-only systems. WLS problem is linear if only phasor measurements of voltage and current are considered. The measurement vector has the following form \cite{AntonioJ.Conejo2018}:
\begin{equation}\label{phasor}
z =  \begin{bmatrix} V^{m} \\ I^{m} \end{bmatrix} = \begin{bmatrix} U \\  Y_b \cdot \Gamma \end{bmatrix} \cdot [V] + e
\end{equation}
where $V^{m}$ and $I^{m}$ are measurements of voltage and current phasors, respectively; $U$ is the identity matrix; $Y_b = [g+jb]$ is the complex network branch admittance matrix; $\Gamma$ is the branch to bus incidence matrix; $V$ is the complex vector of bus voltage phasors; $e$ is the measurement error vector. Equation (\ref{phasor}) consists of complex numbers and can be converted into the rectangular form \cite{AntonioJ.Conejo2018}:
\begin{multline}\label{rectangular}
z =  \begin{bmatrix} V^{m}_r \\ V^{m}_x \\ I^{m}_r \\ I^{m}_x \end{bmatrix} = \begin{bmatrix} U && 0 \\  0 && U \\ g \cdot \Gamma && -b \cdot \Gamma \\ b \cdot \Gamma && g \cdot \Gamma \end{bmatrix} \cdot \begin{bmatrix} V_r \\ V_x \end{bmatrix} + e \\= H_0 \cdot x_0 + e
\end{multline}
where subscripts $r$ and $x$ reflect the real and imaginary components, respectively; $x^{T}_0 = \begin{bmatrix} V_r, V_x \end{bmatrix}$ is the state vector of the system, which is estimated by neglecting all shunts; $H_0$ is the Jacobian which is built by neglecting shunt connections.

The number of measurements should be greater than number of parameters to estimate. Optimal solution for the  WLS model is can be found by close formulation \cite{AntonioJ.Conejo2018}:
\begin{equation} \label{estimate}
\Delta \hat{x} = (H^TR^{-1}H)^{-1}H^TR^{-1}\Delta z
\end{equation}
where $H = H_0 + H_{shunt}$ is the Jacobian matrix accounting for shunt connections; $R$ is the measurement covariance matrix of errors; $\Delta z$ is the measurement residual.
Then, the estimated value of the measurement residual $\Delta z$ is as follows:
\begin{equation}\label{delta_z}
    \Delta \hat{z} = H \Delta \hat{x} = K \Delta z
\end{equation}
where $K = H(H^TR^{-1}H)^{-1}H^TR^{-1}$. \par
Measurement residuals can be expressed in terms of measurement errors \cite{AntonioJ.Conejo2018}:
\begin{multline}\label{residual}
r = \Delta z - \Delta \hat{z} = (U-K)\Delta z \\= (U - K)(H\Delta x+e)= (U-K)e=Se
\end{multline} \par
Here, $S$ is the residual sensitivity matrix, which is expressed as follows \cite{AntonioJ.Conejo2018}:
\begin{equation}\label{sensitivity}
S = U - H(H^TR^{-1}H)^{-1}H^TR^{-1}
\end{equation}

The addition of new PMUs in the system should result in an increased trace of the matrix $S$. Diagonal elements of the matrix $S$ refer to the variance (accuracy) of the residual errors, while the non-diagonal vectors display the covariance between residuals. Thus is, the matrix $S$ maps measurement errors to the residual terms errors. In this work, we consider diagonal elements of the sensitivity matrix $S$. 
%Further, we investigate only the property of diagonal elements of the matrix $S$.  

\section{Proposed Methodology}\label{PropMet}
In power system operation, the placement of PMU sensors is conducted in several stages. One reason for the multi-stage location of sensors can be, for example, a limited annual budget. Thus, the grid operators have an incentive to gain the maximal increase of measurement accuracy for each additionally installed sensor. In this work, we focus on the optimal multi-stage PMU location, which leads to the highest accuracy increase. The decision to install a subset of sensors $A$ at year $i$ is optimal if it leads to the highest increase of measurement accuracy. The presence of submodularity property in the considered problem would mean that a subset of sensors $B$ obtained after making an optimal decision in $j$ consecutive stages is the same as locating all sensors of a subset $B$ at once. In this section, we explore if the residual sensitivity matrix $S$ has a submodularity property and propose our priority list solution. In order to show that the matrix $S$ has a submodularity property, we need to provide solid analytical proof, while to show that $S$ does not a submodularity property finding one counter-example is sufficient.

\subsection{Combinatorial Approach}\label{CombApr}
Multi-stage location of PMUs might be conducted for different purposes. In \cite{Dua2008}, the authors consider a problem of optimal multi-stage placement of PMUs for observability purpose. Let's denote by $\nu$ the subset of PMUs, which is sufficient to make the system observable and has the smallest cardinality. Also, we denote the set containing all buses of the network by $\Omega$. Thus, solutions of PMU location problem for increasing accuracy in the state estimation problem, have the cardinalities between $|\nu|$ and $|\Omega|$. Next, for checking submodularity property, we define two additional subsets $A$ and $B$, which should satisfy:
\begin{equation}\label{ABSets}
\nu \subseteq A \subseteq B \subset \Omega 
\end{equation}

At each stage we add one PMU sensor, so $|s| = 1$. The residual sensitivity matrix $S$ in (\ref{sensitivity}) has a submodular structure, if inequality (\ref{submodular}) holds in all possible combinations of subsets $A$ and $B$. 
%One of the contributions of this work is derived an 
Analytical expression for all possible combinations, denoted by $\alpha$, is equal to:
\begin{equation}\label{NCombinations}
\alpha = \begin{pmatrix} |\Omega| - |\nu| \\ |A| - |\nu| \end{pmatrix} \cdot \begin{pmatrix} |\Omega| - |A| \\ |B| - |A| \end{pmatrix} \cdot \begin{pmatrix} |\Omega| - |B| \\ 1 \end{pmatrix}
\end{equation}
where $\begin{pmatrix} X \\ Y \end{pmatrix}$ is the binomial operator.

We use (\ref{NCombinations}) to verify the number of all combinations during combinatorial brute-force simulations.

% \subsection{Budget-based approach}\label{BudBas}
% formulation via $tr(S)/|S|$

% \subsection{Greedy approach}\label{Greed}
% formulation via $tr(S)/|S|$

\subsection{Submodularity of the Residual Sensitivity Matrix}\label{Metric}
Diagonal elements of the residual sensitivity matrix $S$  relates to the accuracy of the corresponding residual of PMUs in the state estimation problem. Lower values correspond to higher accuracy. We use the following metrics to assess the diagonal elements of the $S$ matrix: minimum and maximum values, sum and average of the diagonal elements. In the next section, we utilize these metrics for PMU location in IEEE 14-bus and 118-bus cases.

\section{Numerical Results}\label{NumRes}
This section defines the most informative metric for analysis of the submodularity property of the matrix $S$. Next, we find several counter-examples proving that the residual sensitivity matrix $S$ does not have a submodular structure in IEEE 14-bus and 118-bus cases. There are bus-type and branch-type PMUs \cite{AntonioJ.Conejo2018}. In this work, we consider bus-type PMUs, which are assumed to have either an unlimited or limited number of inputs for measuring voltage and current phasors at a given bus and on their incident branches. For practical systems, the number of branch inputs is defined as 8 \cite{Meliopoulos2006}. For tested IEEE 14-bus and 118-bus cases, this limit has not been exceeded. Finally, we compare the performance of the optimal budget-constrained and greedy algorithms and provide recommendations for the use of the methods.

\subsection{Defining the Evaluation Metric}\label{DefEvMet}
\begin{figure}
\centering
%   \includestandalone[width=0.3\textwidth]{IEEE14bus}%     without .tex extension
  \includegraphics[width=0.3\textwidth]{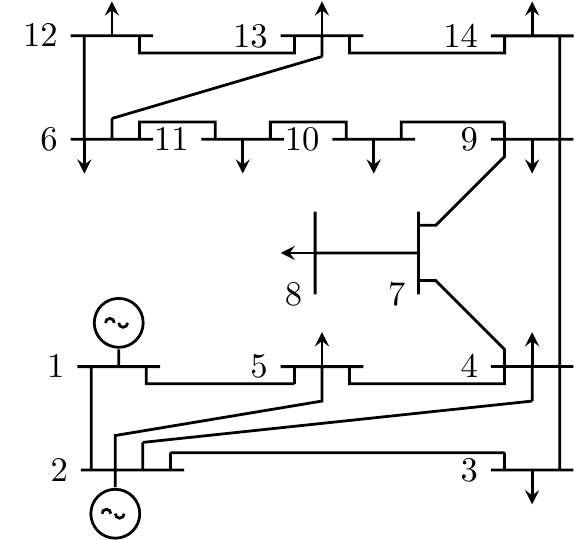}%     without .tex extension
  \caption{IEEE 14-bus system}
  \label{fig:tikz:IEEE_14}
\end{figure}

IEEE 14-bus system is depicted in Fig.\ref{fig:tikz:IEEE_14}. The minimal number of PMU sensors to make the system observable is 4, and the sensors should be installed in nodes 2, 6, 7, 9 \cite{Dua2008}, 
%By the agreed notations in section \ref{CombApr}, we 
denoted $\nu = [2, 6, 7, 9]$. 
Diagonal elements of the $S$ matrix contain valuable information on accuracy. The lower values mean the greater accuracy level. However, the submodularity property (\ref{submodular}) cannot be checked for a vector of diagonal elements. To choose the most informative metric of the matrix $S$, we present the minimum and maximum values, sum and average of its diagonal elements for IEEE 14-bus case in Table \ref{table:metrics14}.

\begin{table}[!h]
\caption{Various metrics for diagonal elements of matrix $S$}
\begin{center}
\begin{tabular}{| c | c | c | c | c |}
\hline
\textbf{Node(s) added } & \multicolumn{4}{ c |}{\textbf{diag(S)}}  \\ 
\cline{2-5}
\textbf{to set $\nu$} & \textbf{min} & \textbf{max} & \textbf{sum} & \textbf{average} \\
\hline
- & 0.5456 & 0.9998 & 28.0000 & 0.7777  \\ \hline
1 & 0.4295 & 0.9992 & 29.9999 & 0.7499  \\ \hline
3 & 0.4303 & 0.9993 &  29.9999 &  0.7499 \\ \hline
4 & 0.5435 &  0.9996 & 32.0000 & 0.7619 \\ \hline
5 & 0.5456 &  0.9997 & 32.0000 & 0.7619 \\ \hline
10 & 0.4221 &  0.9998 & 30.0000 & 0.7500 \\ \hline
14 & 0.4262 &  0.9998 & 29.9999 & 0.7499 \\ \hline
10, 14 & 0.4071 &  0.9998 & 32.0000 & 0.7272 \\ \hline
\end{tabular}
\vspace{-0.4cm}
\end{center}
\label{table:metrics14}
\end{table}
%
%Next, we provide an analysis of 4 metrics in Table \ref{table:metrics14} and choose the most appropriate one. 
Adding a PMU sensor into the system changes the dimension of the residual sensitivity matrix $S$. As a result, the sum of diagonal elements of the matrix $S$ increases with a greater number of PMUs. Thus, the sum poorly displays the change of the residual sensitivity matrix. The minimum and maximum diagonal elements might not be changing while adding additional PMU sensors. For example, the smallest diagonal element is equal to 0.5456, both, for the cases without any additional PMU, and when PMU is added in node 5. Similarly, the biggest diagonal element is equal to 0.9998 for the initial set $\nu$ and when PMU sensors are added in node 10 and/or node 14. As a result, metrics of minimum and maximum of diagonal elements are not informative. Finally, in Table \ref{table:metrics14} we see that the average of diagonal elements consistently decreases when one more PMU is added in the system, so it is used as the main metric and denoted as av(diag(S)) further.

\subsection{Submodularity Check}\label{SubmodCheck}
In this section, we check if the average of diagonal elements in the matrix $S$ has a submodular structure. We conduct our simulations on IEEE 14-bus and 118-bus systems. 

\subsubsection{IEEE 14-bus case}
As mentioned in section \ref{DefEvMet}, the cardinality of the smallest set required to make the system observable is $|\nu| = 4$ and the cardinality of the set containing all nodes is $|\Omega| = 14$. Next, let's choose all subsets $A$ and $B$, which have cardinalities $|A| = 12$, $|B| = 13$ and satisfy $(\ref{ABSets})$. The chosen cardinalities lead to a small number of combinations $\alpha$; thus, decrease the computation burden. According to (\ref{NCombinations}), number of all possible combinations of PMU placement is $\alpha = 90$. As the number of combinations is low, we check each of the cases with a brute-force approach and present results in Table \ref{table: subModCheck}. As it follows from Table \ref{table: subModCheck}, 78 combinations are submodular, and 12 combinations are supermodular, which means that the PMU location problem does not have a distinct submodularity property.
\begin{table}[htb]
\caption{Submodularity check with \textbf{av(diag(S))} metric for IEEE 14-bus and IEEE 118-bus systems}
\resizebox{\linewidth}{!}{  
\begin{tabular}{|l|c|c|c|c|c|}
\hline
 & $\nu$ & $|A|$ & $|B|$ & \textbf{\begin{tabular}[c]{@{}c@{}}Submodular \\ cases\end{tabular}} & \textbf{\begin{tabular}[c]{@{}c@{}}Supermodular \\ cases\end{tabular}} \\ \hline
\textbf{IEEE 14-bus case} & 4 & 12 & 13 & 78 & 12 \\ \hline
\textbf{IEEE 118-bus case} & 37 & 116 & 117 & 6364 & 116 \\ \hline
\end{tabular}}\label{table: subModCheck}
\end{table}

% \begin{table}[!h]
% \caption{Submodularity check with \textbf{av(diag(S))} metric for IEEE 14-bus system with $|A|$ = 12, $|B|$ = 13.}
% \begin{center}
% \begin{tabular}{ | c | c | c | } 
% \hline
% Submodular & 78 \\ 
% \hline
% Supermodular & 12 \\ 
% \hline
% \end{tabular}
% \end{center}
% able:modul14
% \end{table}

\subsubsection{IEEE 118-bus case}
We find that the cardinality of the smallest set required to make the system observable is $|\nu| = 37$, and the cardinality of the set containing all nodes is $|\Omega| = 118$. Next, we consider subsets $|A| = 116$ and $|B| = 117$, which provide a small number of possible combinations. Then, according to (\ref{NCombinations}), number of all possible combinations of PMU placement is $\alpha = 6480$. By automated brute-force approach, we check each of the cases for submodularity property and present results in Table \ref{table: subModCheck}. As it follows from Table \ref{table: subModCheck}, 6364 combinations are submodular, and 116 combinations are supermodular, which means that the system does not have distinct submodularity property.

% \begin{table}[!h]
% \caption{Submodularity check with \textbf{av(diag(S))} metric for IEEE 118-bus system with $|A|$ = 116, $|B|$ = 117.}
% \begin{center}
% \begin{tabular}{ | c | c | c | } 
% \hline
% Submodular & 6364 \\ 
% \hline
% Supermodular & 116 \\ 
% \hline
% \end{tabular}
% \end{center}
% \label{table:modul118}
% \end{table}

Both counterexamples prove that the average of the matrix $S$ diagonal does not have a submodular structure. As a result, setting PMU sensors in $N$ stages provides different results than setting all PMUs at once. However, it is not clear if the budget-constrained algorithm is applicable to the real-world problem of PMU location. In the next section, we compare the budget-constrained and greedy algorithms for multi-stage PMU location and define the most appropriate algorithm.

\subsection{Comparison of optimal budget-constrained and greedy algorithms for multi-stage PMU location problem}
In this section, we compare the optimal budget-constrained and greedy algorithms during multi-stage PMU installation. We use IEEE 14-bus system with pre-installed PMU sensors in buses $\nu = [2, 6, 7, 9]$, which makes it fully observable. We add one PMU sensor at a time following the logic of the budget-constrained and greedy algorithms until PMU sensors are installed in all nodes, i.e., installing 10 PMUs. As discussed in section \ref{NumRes}, we use the average of the diagonal elements of the matrix $S$ as a comparison metric. The numerical results are shown in Table \ref{table:glob_greed}.

\begin{table*}[!t]
\caption{Comparison of optimal budget-constrained and greedy algorithms for multi-stage PMU location problem}
\begin{center}
\begin{tabular}{| c | c | c | c | c |}
\hline
\textbf{Number of additional PMUs/} & \multicolumn{2}{ c |}{\textbf{Budget-constrained algorithm}} & \multicolumn{2}{ c| }{\textbf{Greedy algorithm}} \\ 
\cline{2-5}
\textbf{Stage number} & \textbf{Added node(s)} & \textbf{av(diag(S))} & \textbf{Added node(s)} & \textbf{av(diag(S))} \\
\hline
1 & 8 & 0.7368 & 8 & 0.7368  \\ \hline
2 & 8,14 & 0.7143 &  8,14 &  0.7143 \\ \hline
3 & 8,10,11 &  0.6818 & 8,14,11 & 0.6957 \\ \hline
4 & 1,8,10,11 & 0.6667  & 8,14,11,10 & 0.6667 \\ \hline
5 & 1,8,10,11,14 & 0.6538  & 8,14,11,10,1 & 0.6538 \\ \hline
6 & 8,10,11,12,13,14 & 0.6296  & 8,14,11,10,1,13 & 0.6429 \\ \hline
7 & 1,8,10,11,12,13,14 & 0.6207  & 8,14,11,10,1,13,12 & 0.6207 \\ \hline
8 & 1,5,8,10,11,12,13,14 & 0.6129  & 8,14,11,10,1,13,12,5 & 0.6129 \\ \hline
9 & 1,3,4,8,10,11,12,13,14 & 0.6061  & 8,14,11,10,1,13,12,5,3 & 0.6061 \\ \hline
10 & 1,3,4,5,8,10,11,12,13,14 & 0.5882  & 8,14,11,10,1,13,12,5,3,4 & 0.5882 \\ \hline
\end{tabular}
\end{center}
\label{table:glob_greed}
\end{table*}

For the budget-constrained algorithm, in $i^{th}$ stage, i.e. $i^{th}$ line of Table \ref{table:glob_greed}, the $i$-best candidate nodes for PMU placement are chosen, which provide the lowest resulting value of av(diag(S)) metric. We see that the budget-constrained algorithm does not necessarily keep the solutions of the previous stages in the solution of the next stages. For example, the budget-constrained algorithm at the second stage select nodes 8, 14, but at the third stage select nodes 8, 10, 11, thus eliminating the previously selected node 14 from the third stage. The algorithm ensures that selected nodes result in the smallest av(diag(S)) value for each stage's maximum allowed cardinality.

The greedy algorithm at the first stage follows the same logic as the budget-constrained algorithm. However, beginning from the second stage, it preserves the previously made solutions in future solutions. In Table \ref{table:glob_greed}, the newly added nodes from the greedy algorithm are the last elements of each line. The greedy algorithm ensures that the chosen node at each stage provides the lowest value of av(diag(S)) under the constraint that the previously chosen nodes are present in the current solution.

Comparing the budget-constrained and greedy algorithms, we conclude the following. First, we see in Table \ref{table:glob_greed} that the budget-constrained and greedy algorithm add the same nodes in stages 1, 2, 5, 7, 8, 10. In stages 4 and 9, the greedy algorithm chooses different nodes to add, but the resulting av(diag(S)) value is the same as in the budget-constrained algorithm. In stages 3 and 6, the choice of the greedy algorithm to add different nodes resulted in a higher av(diag(S)) value than in the budget-constrained algorithm. Second, the choices of the budget-constrained algorithm are not consistent with stages, as previously installed PMUs can be removed from the future solutions. At the same time, the greedy algorithm preserves the previously found solutions in all the next decisions. The latter property and the proximity of the av(diag(S)) values to the budget-constrained algorithm make the greedy algorithm the only alternative for practical installation of PMU sensors in the power systems.

\section{Conclusion and Future Work}\label{Conc}
In this work, we propose a decision-making framework for the incremental upgrade of power system components. Our paper has the following contributions. First, we, for the first time, formulate the multi-stage PMU location problem to increase the accuracy in the state estimation problem. Second, we investigate various metrics to analyze the diagonal elements of the residual sensitivity matrix and find that the average provides the most consistent results for upgrading the network. Third, we find counter-examples, which prove that the problem of PMU location for increasing measurement accuracy does not have a submodular structure. Fourth, we numerically compare the performance of the budget-constrained and greedy algorithms for the considered problem and conclude that the greedy algorithm provides results close to the global optimum and is the only practically available option.
Future work includes investigation of the submodularity property in other decision-making problems for an upgrade of power system components in a multi-stage manner and application of the proposed approach in larger real power systems.
\bibliographystyle{IEEEtran}
\bibliography{Priority_list_arXiv}

\end{document}